\documentclass[10pt]{article}%
\usepackage{amsmath}
\usepackage{amsfonts}
\usepackage{amssymb}
\usepackage{graphicx}%
\providecommand{\U}[1]{\protect\rule{.1in}{.1in}}
\providecommand{\U}[1]{\protect\rule{.1in}{.1in}}
\begin{document}

\begin{center}

{\Large \textbf{M-indeterminate distributions in quantum mechanics and the
non-overlapping wave function paradox}}

\bigskip

{\bf R. Sala Mayato$^a$, P. Loughlin$^b$, L. Cohen$^c$}

\bigskip
$^a$ Departamento de F\'isica and IUdEA, Universidad de La Laguna, La Laguna 38203, Tenerife, Spain\\
$^b$ Departments of Bioengineering, and Electrical \& Computer Engineering, University of Pittsburgh, Pittsburgh, PA 15261, USA\\
$^c$ Department of Physics, Hunter College of the City Universiy of New York, 695 Park Ave., New York, NY 10065, USA
\end{center}

\bigskip

\noindent\textbf{{Abstract:}} We consider the non-overlapping wave function paradox of Aharanov \textit{et al.}, wherein the relative phase between two wave functions cannot be measured by the moments of position or momentum.  We show that there is an unlimited number of other expectation values that depend on the phase.  We further show that the Wigner distribution is M-indeterminate, that is, a distribution whose moments do not uniquely determine the distribution.  We generalize to more than two non-overlapping functions.  We consider arbitrary representations and show there is an unlimited number of M-indeterminate distributions.   The dual case of non-overlapping momentum functions is also considered.

\bigskip


\bigskip

\section{Introduction}

In a series of papers Aharanov \textit{et al.} \cite{ahar1,ahar2, ahar3,
ahar-book, toll}, and others \cite{semon-I, semon-II, pope}, discussed a
paradox arising when one considers a wave function that consists of the sum of
two non-overlapping functions. In particular they considered a wave function
of the form%
\begin{equation}
\psi(x)=\frac{1}{\sqrt{2}}\left(  \psi_{1}(x)+e^{i\alpha}\psi_{2}(x)\right)
\label{eq1}%
\end{equation}
where $\alpha$ is real and is the relative phase between the two functions
$\psi_{1}(x)$ and $\psi_{2}(x)$, each of which is normalized to one. The issue
is how to determine the relative phase for the case when $\psi_{1}(x)$ and
$\psi_{2}(x)$ are of finite extent and do not overlap,
\begin{equation}
\psi_{1}^{\ast}(x)\psi_{2}(x)=0
\end{equation}
Accordingly, the position distribution,
\begin{equation}
P(x)=\left\vert \psi(x)\right\vert ^{2}=\frac{1}{2}\left\vert \psi
_{1}(x)+e^{i\alpha}\psi_{2}(x)\right\vert ^{2}\,=\,\frac{1}{2}\left(
\left\vert \psi_{1}(x)\right\vert ^{2}+\left\vert \psi_{2}(x)\right\vert
^{2}\right)
\end{equation}
is independent of $\alpha$ and hence so, too, are the position moments.
Indeed, the expected value of any function of position will be independent of
$\alpha$. Aharanov \textit{et al.} then showed that the momentum moments
\begin{align}
\left\langle p^{n}\right\rangle  &  =\frac{1}{2}\int\left(  \psi_{1}^{\ast
}(x)+e^{-i\alpha}\psi_{2}^{\ast}(x)\right)  \left(  \frac{h}{i}\frac{d}%
{dx}\right)  ^{n}\left(  \psi_{1}(x)+e^{i\alpha}\psi_{2}(x)\right)  dx\\
&  =\frac{1}{2}\int\psi_{1}^{\ast}(x)\left(  \frac{h}{i}\frac{d}{dx}\right)
^{n}\psi_{1}(x)dx+\frac{1}{2}\int\psi_{2}^{\ast}(x)\left(  \frac{h}{i}\frac
{d}{dx}\right)  ^{n}\psi_{2}(x)dx
\label{eq5}%
\end{align}
likewise are independent of $\alpha$. This results in an apparent paradox.
We consider various aspects of the problem in the context of what are known as
\textit{M-indeterminate} distributions, that is, a distribution whose moments
do not uniquely determine the distribution \cite{akhi}.

\section{Momentum probability distribution}

\label{mom-distribution}

We begin with the case of the sum of two non-overlapping wave functions, Eq.
\eqref{eq1}, that are translated versions of each other,
\[
\psi_{1}(x)=\left\{
\begin{array}
[c]{ll}%
f(x)\, & 0\leq x\leq a\\
0,\, & \mathrm{otherwise}%
\end{array}
\right.  ;\qquad\psi_{2}(x)=\left\{
\begin{array}
[c]{ll}%
f(x-L)\, & a<L\leq x\leq L+a\\
0,\, & \mathrm{otherwise}%
\end{array}
\right.
\]
The momentum wave function is
\begin{equation}
\varphi(p)=\frac{1}{\sqrt{2\pi\hbar}}\int\psi(x)\,\,e^{-ipx/\hbar}%
dx\,=\,\frac{1}{\sqrt{2}}\left(  \varphi_{1}(p)+e^{i\alpha}\varphi
_{2}(p)\right)
\end{equation}
where $\varphi_{1}(p)$ and $\varphi_{2}(p)$ are the momentum wave function of
$\psi_{1}(x)$ and $\psi_{2}(x)$ respectively.

Letting
\begin{equation}
F(p)=\varphi_{1}(p)=\frac{1}{\sqrt{2\pi\hbar}}\int_{0}^{a}%
f(x)\,\,e^{-ipx/\hbar}dx
\end{equation}
we have that
\begin{equation}
\varphi(p)=\,\frac{1}{\sqrt{2}}F(p)(1+e^{i(\alpha-pL/\hbar)})
\end{equation}
Hence, the probability distribution of momentum is given by
\begin{equation}
P(p)\,=\,\left\vert \varphi(p)\right\vert ^{2}\,=\,\left\vert F(p)\right\vert
^{2}\left[  1+\cos\left(  pL/\hbar-\alpha\right)  \right]  \label{eq11}%
\end{equation}
Observe that the momentum distribution depends on $\alpha$. However, as
mentioned above the moments $\langle p^{n}\rangle$ do not. Before considering
this apparent paradox further, we give an alternate derivation of the moment
independence on $\alpha$ directly from the distribution, as follows.

Note that
\begin{equation}
\,\left\vert F(p)\right\vert ^{2}=\frac{1}{2\pi\hbar}\left\vert \int_{0}%
^{a}f(x)\,\,e^{-ipx/\hbar}dx\right\vert ^{2}%
\end{equation}
is a proper normalized distribution, meaning $\int\left\vert F(p)\right\vert
^{2}\,dp\,=1$. Thus, we must have
\begin{equation}
\int\left\vert F(p)\right\vert ^{2}\cos\left(  pL/\hbar-\alpha\right)  \,dp=0
\label{aa}%
\end{equation}
Differentiating Eq. (\ref{aa}) with respect to $\alpha$ yields%
\begin{equation}
\int\left\vert F(p)\right\vert ^{2}\sin\left(  pL/\hbar-\alpha\right)  \,dp=0
\label{b}%
\end{equation}
If we further differentiate Eq. (\ref{aa}) and Eq. (\ref{b}) with respect to
$L$, it follows that
\begin{equation}
\int p^{n}\left\vert F(p)\right\vert ^{2}\cos\left(  pL/\hbar-\alpha\right)
\,dp=0
\end{equation}
and therefore, as previously shown by way of Eq. \eqref{eq5},
the moments are independent of $\alpha$ and are given by
\begin{equation}
\left\langle p^{n}\right\rangle =\int p^{n}\left\vert F(p)\right\vert ^{2}\,dp
\end{equation}

Thus the moments of $\left\vert F(p)\right\vert ^{2}$ and $P(p)\,$are
identical, even though $P(p)\neq\left\vert F(p)\right\vert ^{2}$.
Distributions of the form of Eq. \eqref{eq11} are moment-indeterminate (or
\textquotedblleft M-indeterminate\textquotedblright), an observation first
made by Semon and Taylor in the context of the Aharanov-Bohm Effect
\cite{semon-II}. We revisit this distribution and the non-overlapping wave
function paradox from a number of perspectives. First, we explain the
M-indeterminate nature of the distribution in terms of the characteristic
function. This understanding of the root of the paradox allows us to give a
general condition for an unlimited number of expectation values that do depend
on the phase factor. We generalize to more than two non-overlapping functions
and obtain a general expression for a family of M-indeterminate momentum
distributions. We also generalize to arbitrary representations, which allows
the generation of an infinite number of M-indeterminate distributions from the
wave function. Furthermore, we show that while the quantum mechanical current 
does not depend on
the phase factor, its dual, which we call the quantum mechancial group delay, does. 
We show that the Wigner distribution,
among other phase space distributions, is M-indeterminate for non-overlapping
wave functions. We also consider the dual problem, where the momentum wave
function consists of two non-overlapping functions. 

Also, we point out that while the position wave function has no interference term,
the momentum wave function does since it extends over all momentum space. This
must always be the case since the position and momentum wave functions are
Fourier transform pairs, and Fourier transform pairs can not both be of finite extent.

\section{Momentum characteristic function}

\label{momcharfun}

The problem of M-indeterminate distributions  has a long history, with a
majority of the focus being on devising such distributions as well as
determining if a given distribution is M-indeterminate
\cite{akhi,shoh,stoy1,stoy2,gut,pede,klei}. Our interest is in M-indeterminate
distributions in quantum mechanics, which provides unique challenges and
opportunities because of the way probabilities are obtained.

One way to study aspects of the M-indeterminancy problem, which does not seem
to have been as extensively explored, is by way of the characteristic
function. The distribution $P(p)$ and characteristic function $M(\theta)$ are
Fourier transform pairs,
\begin{equation}
M(\theta)=\int e^{i\theta p}P(p)dp=\left\langle e^{i\theta p}\right\rangle
\quad;\quad P(p)\,=\,\frac{1}{2\pi}\int e^{-i\theta p}M(\theta)d\theta
\end{equation}

For our situation we have
\begin{equation}
M(\theta)=\left\langle e^{i\theta p}\right\rangle =\int e^{i\theta
p}\left\vert F(p)\right\vert ^{2}\left[  1+\cos\left(  pL/\hbar-\alpha\right)
\right]  dp \label{eq16}%
\end{equation}
Since $\left\vert F(p)\right\vert ^{2}$ is a probability distribution, we
define its characteristic function by
\begin{equation}
\label{M_F}M_{F}(\theta)=\int\left\vert F(p)\right\vert ^{2}e^{i\theta p}dp
\end{equation}
by which we obtain
\begin{equation}
M_{F}(\theta)=\left\{
\begin{array}
[c]{ll}%
\int_{0}^{a}f(x)f^{\ast}(x-\theta\hbar)dx,\, & -a/\hbar\leq\theta\leq
a/\hbar\\
0,\, & \mathrm{otherwise}%
\end{array}
\right.
\end{equation}
Notice that as a direct consequence of the finite extent of $f(x)$, the
characteristic function $M_{F}(\theta)$ is also of finite extent (it is zero
for $|\theta|>a/\hbar$). Accordingly, by Fourier properties, the momentum
distribution $P(p)$ extends over all $p$.

The characteristic function $M(\theta)$ as given by Eq. {(\ref{eq16})}\ may be
expressed in terms of $M_{F}(\theta)$ as
\begin{equation}
M(\theta)=M_{F}(\theta)+\frac{1}{2}e^{-i\alpha}M_{F}(\theta+L/\hbar)+\frac
{1}{2}e^{i\alpha}M_{F}(\theta-L/\hbar)\label{eq-charfun}%
\end{equation}
Note that, like the momentum distribution, the characteristic function depends
on $\alpha$. Since in general the moments of a distribution can be obtained
from the characteristic function by
\begin{equation}
\langle\,p^{n}\,\rangle\,=\,\left.  \frac{1}{i^{n}}\ {\frac{{{\partial}^{n}}%
}{{{\partial}\theta^{n}}}}\ M(\theta)\ \right\vert _{\theta\,=\,0}%
\end{equation}
we see again that the moments are independent of $\alpha$ since, with $L>a$,
we have that
\begin{equation}
\left.  M_{F}(\theta\pm L/\hbar)\ \right\vert _{\theta\,=\,0}%
=0\label{M_F-shifted}%
\end{equation}
Note that this result highlights the root of the indeterminancy: Eq.
\eqref{eq-charfun} shows that the terms that depend on $\alpha$ are shifted
such that they are not centered about $\theta=0$. Because these terms are
finite extent and the shift is greater than the half-width of the \textquotedblleft
unshifted\textquotedblright\ characteristic function $M_{F}(\theta)$, the
dependence on $\alpha$ is lost when we evaluate the characteristic function at
$\theta=0$ to obtain the moments. Specifically, we have that
\begin{equation}
\,\left.  \frac{1}{i^{n}}\ {\frac{{{\partial}^{n}}}{{{\partial}\theta^{n}}}%
}\ \left\{  e^{-i\alpha}M_{F}(\theta+L/\hbar)+e^{i\alpha}M_{F}(\theta
-L/\hbar)\right\}  \ \right\vert _{\theta\,=\,0}=0
\end{equation}
Therefore, although $P(p) \ne \left|F(p)\right|^2$, they have identical moments
\begin{equation}
\langle\,p^{n}\,\rangle\,=\,\left.  \frac{1}{i^{n}}\ {\frac{{{\partial}^{n}}%
}{{{\partial}\theta^{n}}}}\ M(\theta)\ \right\vert _{\theta\,=\,0}=\,\left.
\frac{1}{i^{n}}\ {\frac{{{\partial}^{n}}}{{{\partial}\theta^{n}}}}%
\ M_{F}(\theta)\ \right\vert _{\theta\,=\,0}%
\label{p-moms-indeterminate-from-charfun}%
\end{equation}
Hence, the family of distributions $P(p)$ (parameterized by $\alpha$) are M-indeterminate.

Historically, the first example of an M-indeterminate distribution
was devised by Stieltjes but perhaps the best known one is the log-normal
distribution,
\begin{equation}
{P}_{\text{LN}}{(x)=}\frac{1}{x\sqrt{2\pi}}{\exp}\left[  -\frac{({\ln}x)^{2}%
}{2}\right]  \qquad0<x<\infty
\end{equation}
for which the moments are
$\left\langle x^{n}\right\rangle =e^{n^{2}/2},$
but they do not uniquely determine the distribution, since
\begin{equation}
P(x)={P}_{\text{LN}}{(x)}\left[  {1+\beta\sin(2\pi\ln x)}\right]
,\qquad-1\leq\beta\leq1
\end{equation}
is a proper probability distribution that has the same moments as
${P}_{\text{LN}}{(x).}$ Thus, the log-normal is M-indeterminate.

There are two well known criteria for the moment indeterminacy problem, one
that deals with the moments directly and the other deals with the
distribution. We consider distributions \ that range from -$\infty$ to
$\infty$ which is called the Hamburg case otherwise it is called the Stieltjes
case. The Carleman condition is that if all the moments, $\left\langle
x^{n}\right\rangle $, of a distribution are finite and if
\begin{equation}
\sum_{n=1}^{\infty}\frac{1}{\left\langle x^{2n}\right\rangle ^{1/2n}}=\infty
\end{equation}
then the distribution having these moments is unique, that is, it is
M-determinate. This is a sufficient but not necessary condition. The other
criterion is the Krein condition: If
\begin{equation}
-\int_{-\infty}^{\infty}\frac{\,\log P(x)}{1+x^{2}}dx<\infty
\end{equation}
then the moments do not determine a unique distribution, that is, it is an
M-indeterminate distribution. Again, this is a sufficient but not necessary condition.

In our case we have the distribution $P_{F}(p)$ given by
\begin{equation}
P_{F}(p)\,=\,\left\vert F(p)\right\vert ^{2}=\frac{1}{2\pi\hbar}\left\vert
\int_{0}^{a}f(x)\,\,e^{-ipx/\hbar}dx\right\vert ^{2}%
\end{equation}
which we have shown is M-indeterminate since%
\begin{equation}
\label{p-moms-indeterminate-from-distribution}\int p^{n}\,\left\vert
F(p)\right\vert ^{2}dp\,=\int p^{n}\,\,\left\vert F(p)\right\vert ^{2}\left[
1+\cos\left(  pL/\hbar-\alpha\right)  \right]  dp
\end{equation}
However even though we have shown that by construction the distribution
$P_{F}(p)$ \ is M-indeterminate, it would be interesting to apply the above
criteria. A challenge is that because $f(x)$ is of finite extent, and hence so
is the characteristic function, the moments $\langle p^{n}\rangle$ may not
exist for all $n$. In particular, if a function $g(x)$ is infinitely
differentiable, such that all of its Fourier (\textit{i.e.}, $p$-) moments exist,
it does not necessarily follow that the function
$f(x)=g(x)u(x)u(a-x)$ is infinitely differentiable because
of the singularities arising from derivatives of the step function $u(x)$ 
($=1$ for $x\ge0$ and zero otherwise). Accordingly, it follows from the
differentiation theorem of the Fourier transform that not all of the moments
$\langle p^{n} \rangle$ of the momentum distribution corresponding to $f(x)$
will exist. In particular, if the $n=N^{th}$ derivative of $f(x)$ contains a
singularity, \textit{i.e.}, a Dirac delta function, then $|F(p)|^{2}
\sim1/p^{2N}$ for $p>>1$.

For all of the moments to exist, we require finite extent functions that are
infinitely differentiable, such as ``bump'' functions. However,
there are many finite extent functions that are not infinitely differentiable.
Hence, there clearly are M-indeterminate distributions that do not have all
finite moments, as given by Eq. \eqref{p-moms-indeterminate-from-distribution}
(equivalently Eq. \eqref{p-moms-indeterminate-from-charfun}) when $f(x)$ is
not infinitely differentiable, yet in such cases, the Carleman condition can
not be used. We consider it an interesting problem to find conditions of
M-indeterminacy when the moments are not all finite but we are not aware of
any results in that regard.

\section{Expectation values that depend on the phase factor}

We now show that, although the momentum moments do not depend on the phase
factor, there is an unlimited number of other expectation values that do. That
this should be so is of course not surprising, given that the momentum
distribution depends on the phase factor, as does the characteristic function,
Eq. {(\ref{eq16})}, which we re-write here as
\begin{align}
M(\theta)  &  =\int\left(  \cos(\theta p)+i\sin(\theta p)\right)  \left\vert
F(p)\right\vert ^{2}\left[  1+\cos\left(  pL/\hbar-\alpha\right)  \right]
dp
 =\left\langle \cos(\theta p)\right\rangle +i\left\langle \sin(\theta
p)\right\rangle
\end{align}
Therefore at least one of the expectation values $\left\langle \cos(\theta
p)\right\rangle $ or $\left\langle \sin(\theta p)\right\rangle $ must depend
on $\alpha$ for some value(s) of $\theta$. This is analogous to the
expectations of the shift operator proposed by Aharanov \textit{et al.} which
in the momentum representation is given by $e^{ipL/\hbar}$ (they further noted
that operators that are \textquotedblleft exponentials of the position and
momentum\textquotedblright\ will also depend on $\alpha$) \cite{toll}. Here,
we give general conditions on functions $g(p)$ so that the expectation value 
$\langle g(p)\rangle$ 
depends on $\alpha$ for the distributions given by Eq. {(\ref{eq11})}. To
achieve this, we require that
\begin{equation}
\int g(p)\left\vert F(p)\right\vert ^{2}\cos\left(  pL/\hbar-\alpha\right)
dp\,\neq0 \label{eq31}%
\end{equation}
Equivalently, in terms of the characteristic function we have, by Eq.
\eqref{eq-charfun} and the multiplication/convolution property of the Fourier
transform,
\begin{equation}
\left.  \left\{  e^{-i\alpha}\int_{-(a+L)/\hbar}^{(a-L)/\hbar}%
G(\theta^{\prime}-\theta)\ M_{F}(\theta^{\prime}+L/\hbar)\ d\theta^{\prime
}+e^{i\alpha}\int_{-(a-L)/\hbar}^{(a+L)/\hbar}G(\theta^{\prime}%
-\theta)\ M_{F}(\theta-L/\hbar)\ d\theta^{\prime}\right\}  \right\vert
_{\theta=0}\neq0
\end{equation}
where the limits of integration follow from the fact that $M_{F}(\theta)$ is
zero for $|\theta|>a/\hbar$, and
\begin{equation}
G(\theta)=\int g(p)e^{-i\theta p}dp
\end{equation}

Thus, our aim is to find functions $G(\theta)$ such that
\begin{equation}
\int_{-(a+L)/\hbar}^{(a-L)/\hbar}G(\theta^{\prime})\ M_{F}%
(\theta^{\prime}+L/\hbar)\ d\theta^{\prime}\neq0 \quad \hbox{and/or} \quad
\int_{-(a-L)/\hbar}^{(a+L)/\hbar}G(\theta^{\prime})\ M_{F}%
(\theta-L/\hbar)\ d\theta^{\prime}\neq0 \label{charfun-condition}%
\end{equation}
This condition can be satisfied by many functions $G(\theta)$ that are
non-zero over the region of integration. For example taking $G(\theta)=1$
results in $g(p)=\delta(p)$ by which it follows that
\begin{equation}
\langle g(p)\rangle=\int\delta(p)\left\vert F(p)\right\vert ^{2}\cos\left(
pL/\hbar-\alpha\right)  dp\,\,=\,\,\left\vert F(0)\right\vert ^{2}\cos\left(
-\alpha\right)
\end{equation}
More generally, the function
\begin{equation}
G(\theta)=\sqrt{\frac{\pi}{\beta}}e^{-\theta^{2}/(4\beta)}%
\end{equation}
gives
\begin{equation}
g(p)=e^{-\beta p^{2}}%
\end{equation}
which will generally satisfy Eq. \eqref{charfun-condition} and hence $\langle e^{-\beta p^{2}}\rangle$ will
depend on $\alpha$.

\section{Quantum mechanical current and group delay}

The quantum mechanical current is also independent of $\alpha$
\cite{ahar-book}. However, as we show here, its dual, which we define
analogous to the current but in terms of the momentum representation, depends
on $\alpha$.

To obtain the current, we express $\psi(x),\psi_{1}(x),$ and $\psi_{2}(x)$ of
Eq. \eqref{eq1} in terms of their respective amplitude and phase,%
\begin{equation}
\label{wave-function-amp-phase}\psi(x)=R(x)e^{iS(x)/\hbar}=\sqrt{\frac{1}{2}%
}\left(  R_{1}(x)e^{iS_{1}(x)/\hbar}+e^{i\alpha}R_{2}(x)e^{iS_{2}(x)/\hbar
}\right)
\end{equation}
Then, the current is
\begin{equation}
j(x)=\frac{\hslash}{2i}\left(  \psi^{\ast}\frac{d\psi}{dx}-\psi\frac
{d\psi^{\ast}}{dx}\right)  =R^{2}(x)S^{\prime}(x)\,
\end{equation}
where we have taken the mass to be equal to one. Following derivations
analogous to \cite{Loughlin-JASA, Loughlin-SPL}, we have two equivalent
expressions for the current%
\begin{align}
j(x)= R^{2}(x)S^{\prime}(x)  &  =R_{1}^{2}S_{1}^{\prime}+R_{2}%
^{2}S_{2}^{\prime}\, +\frac{1}{2}R_{1}R_{2}\,(S_{1}^{\prime}+S_{2}^{\prime
})\cos(S_{1}/\hbar-S_{2}/\hbar-\alpha)\nonumber\\
&  -\frac{\hbar}{2}(R_{1}R_{2}^{\prime}\,-\,R_{1}^{\prime}R_{2})\sin(S_{1}%
/\hbar-S_{2}/\hbar-\alpha) \label{qm-current}%
\end{align}
and%
\begin{align}
j(x)  &  =\frac{1}{2}\left(  S_{1}^{\prime}+S_{2}^{\prime}\right)  R^{2}%
+\frac{1}{2}\left(  S_{1}^{\prime}-S_{2}^{\prime}\right)  (R_{1}^{2}-R_{2}%
^{2}) \,-\,\hbar\,(R_{1}R_{2}^{\prime}-R_{1}^{\prime}R_{2})\sin\left(  S_{1}%
/\hbar-S_{2}/\hbar-\alpha\right)
\end{align}
where
\begin{equation}
R^{2}=R_{1}^{2}\,+R_{2}^{2}\,+\,2R_{1}\,R_{2}\,\cos(S_{1}/\hbar-S_{2}%
/\hbar-\alpha) \label{R-squared}
\end{equation}

Because the functions do not overlap, we have that $R_{1}R_{2}=0$ and hence
the current is independent of $\alpha$ and is given by
\begin{equation}
j(x)= R_{1}^{2}S_{1}^{\prime}+R_{2}^{2}S_{2}^{\prime}
\end{equation}
This result generalizes as well for $N>2$ non-overlapping functions
\cite{Loughlin-SPL}; that is, the current is a weighted sum of the individual
currents of each wave function, and is independent of any constant phase terms.

We also note that one can think of quantum mechanical current as the local
expectation value of momentum, namely
\begin{equation}
\langle\,\ p\ \,\rangle_{x}=\int p\ W_{\psi}(x,p)dp\label{Wigner-current}%
\end{equation}
where $W_{\psi}(x,p)\ $is the Wigner distribution of the wave function
$\psi(x)$,
\begin{equation}
W_{\psi}(x,p)\,=\,{\frac{1}{2\pi}}\int\psi{^{\ast}}(x-{{{\tfrac{\hslash}{2}}}%
}\tau)\,\psi(x+{{{\tfrac{\hslash}{2}}}}\tau)\,e^{-i\tau p}\,d\tau
\end{equation}
In this case, Eq. \eqref{Wigner-current} equals the current,
\begin{equation}
\langle\,\ p\ \,\rangle_{x}=j(x)= R_{1}^{2}S_{1}^{\prime}+R_{2}^{2}
S_{2}^{\prime}
\end{equation}
We point out that many other quasi-distributions \cite{cohen-book} also yield
this result.

\noindent\textbf{Quantum group delay}. Analogous to group delay in pulse
propagation, we define the \textit{quantum group delay} in terms of the
momentum representation as the dual to quantum mechanical current. Writing the
momentum wave function in terms of its amplitude and phase,
\begin{equation}
\varphi(p)=\,B(p)e^{i\eta(p)/\hbar}\,
\end{equation}
we define the quantum group delay, $\tau(p),$ as
\begin{equation}
\tau(p)=\frac{\hslash}{2i}\left(  \varphi^{\ast}\frac{d\varphi}{dp}%
-\varphi\frac{d\varphi^{\ast}}{dp}\right)  =B^{2}(p)\eta^{\prime}(p)\,
\end{equation}
For the case where we have two arbitrary wave functions with a relative
constant phase shift (Eq. \eqref{wave-function-amp-phase}), the momentum wave
function in terms of amplitude and phase is
\begin{equation}
\varphi(p)=\,B(p)e^{i\eta(p)/\hbar}\,=\,\sqrt{\frac{1}{2}}\left(
B_{1}(p)e^{i\eta_{1}(p)/\hbar}+e^{i\alpha}B_{2}(p)e^{i\eta_{2}(p)/\hbar
}\right)
\end{equation}
Accordingly, the quantum group delay is
\begin{equation} \label{eqA}
\tau(p)=B^{2}\eta^{\prime}=\frac{1}{2}B^{2}\left(  \eta_{1}^{\prime}+\eta
_{2}^{\prime}\right)  +\frac{1}{2}\left(  \eta_{1}^{\prime}-\eta_{2}^{\prime
}\right)  (B_{1}^{2}-B_{2}^{2})\,-\,\hbar\,(B_{1}B_{2}^{\prime}-B_{1}^{\prime}%
B_{2})\sin\left(  \eta_{1}/\hbar-\eta_{2}/\hbar-\alpha\right)
\end{equation}
where
\begin{equation}
B^{2}=B_{1}^{2}+B_{2}^{2}+2B_{1}B_{2}\,\cos\left(  \eta_{1}/\hbar-\eta
_{2}/\hbar-\alpha\right)  \label{eqB}%
\end{equation}

Note that, unlike with the expression for the current, here we have that
$B_{1}(p)B_{2}(p)\neq0$ in general, even if $\psi_{1}(x)\psi_{2}(x)=0$. Hence,
in general, the quantum mechanical group delay depends on the relative phase,
$\alpha$.

For the case of two non-overlapping wave functions where one is a translated
version of the other as considered in Sctn. \ref{mom-distribution}, we have
\begin{align}
B_{2}(p)  &  =B_{1}(p)\ ; \qquad\eta_{2}(p) =\eta_{1}(p)-pL/\hbar
\end{align}
by which it follows that
\begin{align}
\tau(p)  &  =\frac{1}{2}\left(  2\eta_{1}^{\prime}-L/\hbar\right)  \left\vert
F(p)\right\vert ^{2}(1+\cos\left(  pL/\hbar-\alpha\right)  ) \, =\, \left(
2\eta_{1}^{\prime}-L/\hbar\right)  \left\vert F(p)\right\vert ^{2}\cos
^{2}\left(  \frac{pL/\hbar-\alpha}{2}\right)
\end{align}

Analogous to the quantum mechanical current, the group delay can be obtained
as the local expectation value of position from the Wigner distribution,
\begin{equation}
\langle\ x\ \rangle_{p}=\int x\ W_{\psi}(x,p)\ dx=\tau(p)
\end{equation}

\section{M-indeterminate quantum phase space distributions}

The above considerations lead us to determine whether or not the Wigner 
distribution of two non-overlapping wave functions is M-indeterminate. In
particular, are the mixed moments $\langle x^{n}p^{m}\rangle$ of the Wigner distribution,
\begin{equation}
\langle x^{n}p^{m}\rangle=\iint x^{n}p^{m}\ W_{\psi}(x,p)\ dxdp \label{w1}%
\end{equation}
independent of $\alpha$?
For the wave function given by Eq. \eqref{eq1}, the Wigner distribution is
\begin{align}
W_{\psi}(x,p)\,  &  ={{{\tfrac{1}{2}}}}W_{\psi_{1}}(x,p)\,+{{{\tfrac{1}{2}}}%
}W_{\psi_{2}}(x,p)\,+e^{-i\alpha}W_{12}(x,p)\,+e^{i\alpha}W_{21}%
(x,p)\nonumber\\
&  ={{{\tfrac{1}{2}}}}W_{\psi_{1}}(x,p)\,+{{{\tfrac{1}{2}}}}W_{\psi_{2}%
}(x,p)\,+2\cos\alpha\operatorname{Re}\left\{  W_{12}\right\}  +2\sin
\alpha\operatorname{Im}\left\{  W_{12}\right\}  \, \label{w6}%
\end{align}
where $W_{12}$ is the cross Wigner distribution of the functions $\psi_{1}$
and $\psi_{2}$,
\begin{equation}
W_{12}(x,p)\,=\,{\frac{1}{2\pi}}\int\psi_{1}{^{\ast}}(x-{{{\tfrac{\hslash}{2}%
}}}\tau)\,\psi_{2}(x+{{{\tfrac{\hslash}{2}}}}\tau)\,e^{-i\tau p}%
\,d\tau\label{w4}%
\end{equation}
and similarity for $W_{21}$. Thus, the Wigner distribution depends on the
phase factor $\alpha$.

To examine whether the moments as given by Eq. {(\ref{w1})}\ are dependent on
$\alpha$ we note that the first two terms in Eq. {(\ref{w6})} do not depend on
$\alpha$ and hence we have to examine whether
\begin{equation}
\langle x^{n}p^{m}\rangle_{12}=\iint x^{n}p^{m}\ W_{12}(x,p)\,\ dxdp
\label{w7}%
\end{equation}
is zero or not. Substituting Eq. \eqref{w4} into Eq. {(\ref{w7})} gives
\begin{align}
\langle x^{n}p^{m}\rangle_{12}  
&  ={\frac{\,i^{m}}{2\pi}}%
{\displaystyle\iiint}
x^{n}\ \int\psi_{1}{^{\ast}}(x-{{{\tfrac{\hslash}{2}}}}\tau)\,\psi
_{2}(x+{{{\tfrac{\hslash}{2}}}}\tau)\,\frac{\partial^{m}}{\partial\tau^{m}%
}e^{-i\tau p}\,d\tau\ dxdp
\end{align}
and straightforward integration by parts yields
\begin{equation}
\langle x^{n}p^{m}\rangle_{12}=\,i^{m}\sum_{k=0}^{m}\,{\binom{n}{k}}\left(
-{1}\right)  ^{k}\int x^{n}\ \left(  \frac{\partial^{k}}{\partial x^{k}}%
\psi_{1}{^{\ast}}(x)\right)  \left(  \frac{\partial^{m-k}}{\partial x^{m-k}%
}\,\psi_{2}(x)\right)  \ dx \label{w10}%
\end{equation}
which could have also been obtained using the Weyl-McCoy correspondence for
$x^{n}p^{m}$ \cite{cohen-book}. Since the wave functions do not overlap we
have that
\begin{equation}
\langle x^{n}p^{m}\rangle_{12}=0
\end{equation}
This shows that the mixed moments are independent of $\alpha$ and hence the
Wigner distribution is M-indeterminate. Indeed, there is an unlimited number
of M-indeterminate phase space distributions; in particular, if their mixed
moments can be written in terms of a (finite) sum of the form
\begin{equation}
\int\psi^{\ast}(x)\mathbf{x}^{n}\mathbf{p}^{m}\psi(x)dx \label{eq12}%
\end{equation}
then the distribution will be M-indeterminate.

\section{Multiple non-overlapping wave functions}

For the case of more than two non-overlapping wave functions we define
\begin{equation}
\psi(x)=\frac{1}{\sqrt{N}}\sum_{n=1}^{N}\psi_{n}(x)
\label{PL-N-non-overlapping-02a-eq1}%
\end{equation}
and take the individual wave functions to be non-overlapping,
\begin{equation}
\psi_{n}^{\ast}(x)\psi_{m}(x)=0,\quad n\neq m
\label{PL-N-non-overlapping-02a-eq3}%
\end{equation}
Let
\begin{equation}
\psi_{n}(x)=e^{i\alpha_{n}}f\left(  x-nL\right)
\label{PL-N-non-overlapping-02a-eq4}%
\end{equation}
As with the $N=2$ case, the position moments are independent of the phases
$\alpha_{n}$, since the probability distribution of position is given by
\begin{equation}
\left\vert \psi(x)\right\vert ^{2}=\left\vert \frac{1}{\sqrt{N}}\sum_{n=1}%
^{N}\psi_{n}(x)\right\vert ^{2}=\frac{1}{N}\sum_{n=1}^{N}\left\vert \psi
_{n}(x)\right\vert ^{2}%
\end{equation}

The momentum wave function is 
\begin{equation}
\varphi(p)=\frac{1}{\sqrt{2\pi \hbar}}\int\psi(x)e^{-ixp/\hbar}\,dx=\,\frac{F(p)}{\sqrt
{N}}\sum_{n=1}^{N}e^{i(\alpha_{n}-npL/\hbar)}%
\end{equation}
The summation depends on the specific $\alpha_{n}$ and can not in general be
further simplified, but clearly the momentum distribution $|\varphi(p)|^{2}$
depends on the phases $\alpha_{n}$. If we make the simplifying assumption that
{${\alpha_{n}=n\,\alpha}$}${{,}}$ which imposes a constant relative phase
difference between the adjacent $\psi_{n}(x),$ then we have
\begin{equation}
\varphi(p)=\,\frac{F(p)}{\sqrt{N}}\sum_{n=1}^{N}e^{in(\alpha-pL/\hbar)}%
\,=\,F(p)\,e^{i(\alpha-pL/\hbar)/2}\,\frac{\sin{(N(\alpha-pL/\hbar)/2)}}{\sin
{((\alpha-pL/\hbar)/2)}}%
\end{equation}
Therefore, the momentum distribution is 
\begin{equation}
\left\vert \varphi(p)\right\vert ^{2}\,=\,\frac{\left\vert \,F(p)\right\vert
^{2}}{N}\,\sum_{n=1}^{N}\sum_{k=1}^{N}e^{i\alpha(n-k)}\,e^{-ipL(n-k)/\hbar}\,=\,
\frac{\left\vert \,F(p)\right\vert ^{2}}{N}\,\left\vert \frac{\sin
{(N(\alpha-pL/\hbar)/2)}}{\sin{((\alpha-pL/\hbar)/2)}}\right\vert ^{2} \label{eq88}%
\end{equation}
One can show that Eq. {(\ref{eq88})}\ reduces to eq. {(\ref{eq11})}\ for
$N=2.$ The function $\frac{\sin(Nx/2)}{\sin(x/2)}$ appears frequently in
sonar, radar, optics and digital image processing, such as for example a
grating (or line array) of $N$ equi-spaced apertures. It goes by different
names, including the \textquotedblleft periodic sinc,\textquotedblright%
\ \textquotedblleft aliased sinc,\textquotedblright\ \textquotedblleft
circular sinc\textquotedblright\ and \textquotedblleft Dirichlet
function,\textquotedblright although this latter term is used to refer to
other functions as well.

We now show that the momentum distribution for $N$ non-overlapping shifted
functions, Eq. {(\ref{eq88}), } is M-indeterminate by showing that its
moments
\begin{equation}
\left\langle p^{n}\right\rangle =\int p^{n}\left\vert \varphi(p)\right\vert
^{2}\,dp
\end{equation}
are independent of $\alpha$. We approach the issue by way of the
characteristic function, which is given by 
\begin{align}
M(\theta)  &  =\left\langle e^{i\theta p}\right\rangle =\int e^{i\theta
p}\frac{\left\vert \,F(p)\right\vert ^{2}}{N}\,\sum_{n=1}^{N}\sum_{k=1}%
^{N}e^{i\alpha(n-k)}\,e^{-ipL(n-k)/\hbar}dp\\
&  =\frac{1}{N}\,\sum_{n=1}^{N}\sum_{k=1}^{N}e^{i\alpha(n-k)}\,M_{F}%
(\theta-L(n-k)/\hbar)
\end{align}
where $M_{F}$ is the characteristic function of $\left\vert \,F(p)\right\vert
^{2}$ as given by Eq. {(\ref{M_F})}. Extracting the $n=k$ term we have
\begin{align}
\,M(\theta)  &  =M_{F}(\theta)+\frac{1}{N}\,\sum_{n,k=1;n\neq k}^{N}%
e^{i\alpha(n-k)}\,M_{F}(\theta-L(n-k)/\hbar)\\
&  =M_{F}(\theta)+\frac{1}{N}\,\sum_{n>k}^{N}e^{i\alpha(n-k)}\,M_{F}%
(\theta-L(n-k)/\hbar)+\frac{1}{N}\,\sum_{n<k}^{N}e^{-i\alpha(n-k)}\,M_{F}%
(\theta+L(n-k)/\hbar)
\end{align}
Analogous to the situation in Sctn. \ref{momcharfun}, we have that $\left.
M_{F}(\theta\pm L(n-k)/\hbar)\right\vert _{\theta=0}=0$ and hence the moments are
given by
\begin{equation}
\langle\,p^{n}\,\rangle\,=\,\left.  \frac{1}{i^{n}}\ {\frac{{{\partial}^{n}}%
}{{{\partial}\theta^{n}}}}\ M(\theta)\ \right\vert _{\theta\,=\,0}=\,\left.
\frac{1}{i^{n}}\ {\frac{{{\partial}^{n}}}{{{\partial}\theta^{n}}}}%
\ M_{F}(\theta)\ \right\vert _{\theta\,=\,0}%
\end{equation}
and therefore the momentum distribution for $N$ non-overlapping shifted functions
is M-indeterminate.

\section{Probability of other observables}

We consider here the transformation of the wave function given by Eq.
{(\ref{eq1})}\ to a general representation, and whether or not the
corresponding distribution is M-indeterminate.

For an observable represented by the Hermitian operator $\mathbf{A}$ the
eigenvalue problem (we consider the discrete case),
\begin{equation}
\mathbf{A}\,u_{n}(x)\,=\,a_{n}\,u_{n}(x)
\end{equation}
results in real eigenvalues, $\,a_{n}$, which are the measurable quantities.
The eigenfunctions,$\,u_{n}(x),$ are complete and orthogonal
\begin{equation}
\int\,u_{k}^{\ast}(x)\,u_{n}(x)\,dx\,=\,\delta_{kn}\ ; \qquad
{\displaystyle\sum\limits_{n}} \,u_{n}^{\ast}(x^{\prime})\,u_{n}%
(x)\,=\,\delta(x-x^{\prime})\
\end{equation}
Expanding the wave function, $\psi(x),$ as
\begin{equation}
\psi(x)=%
{\displaystyle\sum\limits_{n}}
\,c_{n}\,u_{n}(x)
\end{equation}
gives the probability, $P(a_{n}),$ of measuring $a_{n},$
\begin{equation}
P(a_{n})=\left\vert c_{n}\right\vert ^{2}=\left\vert \int\psi(x)u_{n}^{\ast
}(x)dx\right\vert ^{2}%
\end{equation}
with
\begin{equation}
c_{n}=\int\psi(x)u_{n}^{\ast}(x)dx
\end{equation}

For the wave function given by Eq. {(\ref{eq1})}, we have
\begin{equation}
c_{n}=\frac{1}{\sqrt{2}}(c_{n}^{(1)}+e^{i\alpha}c_{n}^{(2)})
\end{equation}
where
\begin{equation}
c_{n}^{(1)}=\int\psi_{1}(x)u_{n}^{\ast}(x)dx \ ; \qquad c_{n}^{(2)}=\int
\psi_{2}(x)u_{n}^{\ast}(x)dx
\end{equation}
Therefore%
\begin{equation}
P(a_{n})=\left\vert \int\psi(x)u_{n}^{\ast}(x)dx\right\vert ^{2}=\frac{1}{2}%
{\displaystyle\iint}
\left(  \psi_{1}^{\ast}(x^{\prime})+e^{-i\alpha}\psi_{2}^{\ast}(x^{\prime
})\right)  u_{n}(x^{\prime})\left(  \psi_{1}(x)+e^{i\alpha}\psi_{2}(x)\right)
u_{n}^{\ast}(x)dxdx^{\prime}%
\end{equation}
which gives%
\begin{equation}
P(a_{n})=\frac{1}{2}(P_{1}(a_{n})+P_{2}(a_{n}))+\frac{1}{2}\left(  e^{i\alpha
}c_{n}^{\ast(1)}c_{n}^{(2)}+e^{-i\alpha}c_{n}^{\ast(2)}c_{n}^{(1)}\right)
\end{equation}
where%
\begin{equation}
P_{1}(a_{n})=\left\vert c_{n}^{(1)}\right\vert ^{2}\hspace{0.2in}%
;\hspace{0.2in}P_{2}(a_{n})=\left\vert c_{n}^{(2)}\right\vert ^{2}%
\end{equation}
Now generally speaking, while $\psi_{1}^{\ast}(x)\psi_{2}(x)=0,$ we have
\begin{equation}
c_{n}^{\ast(1)}c_{n}^{(2)}\neq0
\end{equation}
and hence the probability distribution will depend on the phase factor.

Now consider the moments,
\begin{align}
\left\langle \mathbf{A}^{n}\right\rangle =%
{\displaystyle\sum\limits_{n=0}^{\infty}}
a^{n}\left\vert c_{n}\right\vert ^{2}  &  =\frac{1}{2}\int\left(  \psi
_{1}^{\ast}(x)+e^{-i\alpha}\psi_{2}^{\ast}(x)\right)  \mathbf{A}^{n}\left(
\psi_{1}(x)+e^{i\alpha}\psi_{2}(x)\right)  dx\\
&  =\frac{1}{2}\left\langle \mathbf{A}^{n}\right\rangle _{1}+\frac{1}%
{2}\left\langle \mathbf{A}^{n}\right\rangle _{2}+2\operatorname{Re}\left(
e^{-i\alpha}\int\psi_{2}^{\ast}(x)\mathbf{A}^{n}\psi_{1}(x)dx\right) \label{9b}
\end{align}
where $\left\langle \mathbf{A}^{n}\right\rangle _{1}$ and $\left\langle
\mathbf{A}^{n}\right\rangle _{2}$ are the expectation values taken with the
wave functions $\psi_{1}(x)$ and $\psi_{2}(x)$, respectively, and they do not
depend on $\alpha$. The question then is, are there Hermitian operators such
that $\int\psi_{2}^{\ast}(x)\mathbf{A}^{n}\psi_{1}(x)dx\neq0$? If
$\mathbf{A}^{n}\psi_{1}(x)$ has the same support as $\psi_{1}(x),$ that is, it
is zero over the same interval as $\psi_{1}(x)$, then the integral will be
zero. So, for example if $\mathbf{A}$ is the sum of finite polynomials in
position and momentum then the last term in Eq. (\ref{9b}) will be zero and
hence the moments will be independent of the phase factor. Therefore, for
non-overlapping wave functions, there are many observables that result in
M-indeterminate distributions. Generalization to continuous representations
and distributions is straightforward. We note that in general there will be
interference terms in these representations even though $\psi_{1}^{\ast
}(x)\psi_{2}(x)=0.$

An interesting aspect of the above consideration is that it generates an
infinite number of M-indeterminate distributions, both continuous and
discrete. That is achieved by choosing representations that are generated by a
Hermitian operator. This will be developed in a future paper.

\section{Dual Case}

Finally, we briefly consider the dual case, namely a momentum wave function
that consists of the sum of two non-overlapping momentum wave functions,
\begin{equation}
\varphi(p)=\frac{1}{\sqrt{2}}\left(  \varphi_{1}(p)+e^{i\alpha}\varphi
_{2}(p)\right)
\end{equation}
where $\alpha$ is the relative phase between the two momentum wave functions
$\varphi_{1}(p)$ and $\varphi_{2}(p)$, each of which is normalized to one and
where $\varphi_{1}(p)$ $\ $and $\varphi_{2}(p)$ are of finite extent
\begin{align}
\varphi_{1}(p)  &  =\left\{
\begin{array}
[c]{ll}%
h(p)\, & 0\leq p\leq b\\
0,\, & \mathrm{otherwise}%
\end{array}
\right.  ; \qquad\varphi_{2}(p) =\left\{
\begin{array}
[c]{ll}%
h(p-L)\, & b<L\leq p\leq L+b\\
0,\, & \mathrm{otherwise}%
\end{array}
\right.
\end{align}
such that
\begin{equation}
\varphi_{1}^{\ast}(p)\varphi_{2}(p)=0
\end{equation}
For this case, the momentum distribution, $P(p),$ is independent of $\alpha$,
\begin{equation}
P(p)=\left\vert \varphi(p)\right\vert ^{2}=\frac{1}{2}\left\vert \varphi
_{1}(p)+e^{i\alpha}\varphi_{2}(p)\right\vert ^{2}\,=\,\frac{1}{2}\left(
\left\vert \varphi_{1}(p)\right\vert ^{2}+\left\vert \varphi_{2}(p)\right\vert
^{2}\right)
\end{equation}
whereas the position distribution is not,
\begin{equation}
P(x)\,=\,\left\vert \psi(x)\right\vert ^{2}\,=\,\left\vert H(x)\right\vert
^{2}\left[  1+\cos\left(  xL/\hbar-\alpha\right)  \right]
\end{equation}
where
\begin{equation}
H(x)=\frac{1}{\sqrt{2\pi\hbar}}\int_{0}^{a}h(p)\,\,e^{ipx/\hbar}dp
\end{equation}
Hence, analogous to the previous case, the position and momentum moments are
again independent of $\alpha$, but now it is the position distribution that is M-indeterminate.
An example of this case is considered in \cite{wang-scully}.

It follows that all of the previous results apply here, after transcribing $x$
for $p$ and vice versa. In other words, where in the previous case the quantum
mechanical current was independent of $\alpha$ but the quantum group delay was
not, here the situation is reversed: the quantum mechanical current depends on
$\alpha$ while the quantum group delay does not. Explicitly, for the
calculation of the current we write
\begin{equation}
\psi(x)=\,R(x)e^{iS(x)/\hbar}\,=\,\frac{1}{\sqrt{2}}\left(  R_{1}%
(x)e^{iS_{1}(x)/\hbar}+e^{i\alpha}R_{2}(x)e^{iS_{2}(x)/\hbar)}\right)
\end{equation}
for which the current is given by Eq. \eqref{qm-current}.
However, unlike the previous case, here $R_{1}R_{2}\neq0$, and hence the current
depends on $\alpha$. As a special case, for
\begin{equation}
\varphi_{2}(x)=\varphi_{1}(x-L)
\end{equation}
the current is
\begin{equation}
j(x)=\,S_{1}(x)-R^{2}L/2
\end{equation}
where $R^2$ depends on $\alpha$ and is given by Eq. \eqref{R-squared}.

\bigskip

\section{Conclusion}

\bigskip

We considered the non-overlapping wave function paradox in quantum mechanics,
wherein expectation values of position and momentum are independent of the
relative phase between the wave functions, from the perspective of
M-indeterminate distributions. We showed that, not only do non-overlapping
wave functions with a relative phase difference give rise to M-indeterminate
momentum distributions, but also there is an infinite number of
M-indeterminate distributions, each associated with a different physical
observable. 

What is particularly interesting about the non-overlapping wave function
paradox is the way in which moments and probability distributions are
calculated in quantum mechanics. We showed that the characteristic function
approach is particularly powerful for addressing the non-overlapping wave
function paradox and highlights precisely why the moments are independent of
the phase difference.

With this insight afforded by the characteristic function approach, we
obtained general conditions for other expectation values that depend on the
relative phase. We also defined the quantum mechanical group delay as the dual
to the quantum mechanical current, and showed that the group delay is a
function of the phase difference even though the current is not. This result
is a direct consequence of the fact that, while the non-overlapping wave
functions do not interfere in position space, they do interfere in momentum space.

Considerations of the current and group delay lead naturally to phase space
quasi-distributions, such as the Wigner distribution. In particular, the group
delay and current are local expectation values of the Wigner (and many other)
distributions. We showed that non-overlapping wave functions have
M-indeterminate phase space distributions. A particularly interesting aspect
of the problem here is that phase space distributions may be negative, or even
complex, which is why they are often called \textquotedblleft
quasi-distributions.\textquotedblright\ As such, the usual tests for
M-indeterminate distributions may not apply. We consider it an interesting
future avenue of exploration to study these situations.

We also considered the dual problem, namely non-overlapping wave functions in
momentum space, and showed that all of our results pertain to this case with a
simple transcription of variables and quantities. For non-overlapping momentum
functions, the quantum mechanical current will depend on the relative phase
difference, as will the position distribution but not its moments. For this
case it is the position distribution, not the momentum distribution, that is M-indeterminate.

\section{Acknowledgements}

R. Sala Mayato acknowledges funding by the Spanish MINECO (FIS2017-82855-P, FIS2013-41352-P).


\begin{thebibliography}{99}                                                                                               %


\bibitem {ahar1}Aharonov, Y., Pendleton, H. and Petersen, A., ``Modular Variables in Quantum Theory,'' \textit{Int. J. Th. Phys.} 2(3), 213-230, 1969.

\bibitem {ahar2}Aharonov, Y., Pendleton, H. and Petersen, A., \textquotedblleft
Deterministic Quantum Interference Experiments,\textquotedblright%
\ \textit{Int. J. Th. Phys.} 3(6), 443-442, 1970.

\bibitem {ahar3}Aharonov, Y., in Proc. Int. Symp. Foundations of Quantum
Mechanics and their Technical Implications (eds. Kamefuchi, S. et al.) 10--19, 1984.

\bibitem {ahar-book}Aharonov, Y. and Rohrlich, D., \textit{Quantum Paradoxes},
Wiley-VCH, 2005.

\bibitem {toll}Tollaksen, J., Aharonov, Y., Casher, A., Kaufherr, T., and
Nussinov, S., ``Quantum interference experiments, modular variables and weak
measurements,'' \textit{New J. Phys.} 12, 013023, 2010.

\bibitem {semon-I}Semon, M. and Taylor, J., ``Expectation Values in the
Aharonov-Bohm Effect,'' \textit{Il Nuovo Cimento} 97B(1), 1987.

\bibitem {semon-II}Semon, M. and Taylor, J., ``Expectation Values in the
Aharonov-Bohm Effect -II,'' \textit{Il Nuovo Cimento} 100B(3), 1987.

\bibitem {pope}Popescu, S., \textquotedblleft
Dynamical quantum non-locality,\textquotedblright%
 \textit{Nature Physics} 6, 151-153, 2010.

\bibitem {akhi}Akhiezer, N.I., \textit{The classical moment problem and some
related questions in analysis}, Hafner Publishing Co., New York, 1965.

\bibitem {shoh} Shohat, J.A. and Tamarkin, J.D.,  \textit{The Problem of Moments,}
American Mathematical Society Mathematical Surveys, vol. I. American
Mathematical Society, New York, 1943.

\bibitem {stoy1}Stoyanov, J., \textquotedblleft Determinacy of Distributions by
Their Moments\textquotedblright, In: International Conference on Mathematical
and Statistical Modeling in Honor of Enrique Castillo. June 28-30, 2006

\bibitem {stoy2}Stoyanov, J., \textquotedblleft Krein condition in
probabilistic moment problems\textquotedblright, \textit{Bernoulli} 6, 939-949, 2000.

\bibitem {gut}Gut, A., \textquotedblleft On the moment problem\textquotedblright, \textit{Bernoulli} 8, 407-421, 2002.

\bibitem {pede}Pedersen, H. L., \textquotedblleft On Krein's theorem for
indeterminacy of the classical moment problem\textquotedblright,  \textit{J. Approx.
Theory} 95, 90-100, 1998.

\bibitem {klei}Kleiber, C., \textquotedblleft The generalized lognormal
distribution and the stieltjes moment problem\textquotedblright, \textit{Journal of
Theoretical Probability} 27, 1167-1177, 2014.

\bibitem {Loughlin-JASA}Loughlin, P., ``Spectrographic measurement of
instantaneous frequency and the time-dependent weighted average instantaneous
frequency,'' \textit{The Journal of the Acoustical Society of America}, 105
(1), 264-274, 1999.

\bibitem {Loughlin-SPL}Nho, W. and Loughlin, P., \textquotedblleft When is
instantaneous frequency the average frequency at each time?,\textquotedblright%
\ \textit{IEEE Signal Processing Letters}, 6 (4), 78-80, 1999.

\bibitem {cohen-book}Cohen, L., \textit{Time-Frequency Analysis},
Prentice-Hall, 1995.

\bibitem{wang-scully} Wang, D-W.  and Scully, M.O.,  ``Heisenberg Limit Superradiant Superresolving Metrology,'' {\it Physical Review Letters}, 113, 083601, 2014.



\end{thebibliography}
\end{document}